\documentclass[]{spie}  

\usepackage{amsmath,amsfonts,amssymb}
\usepackage{graphicx}
\usepackage[colorlinks=true, allcolors=blue]{hyperref}
\usepackage{booktabs}
\usepackage{multicol,multirow,makecell}

\title{Performance and Non-adversarial Robustness of the Segment Anything Model 2 in Surgical Video Segmentation}

\author[a,*]{Yiqing Shen}
\author[a,*]{Hao Ding}
\author[a,*]{Xinyuan Shao}
\author[a]{Mathias Unberath}
\affil[a]{Department of Computer Science, Johns Hopkins University, Baltimore, USA}

\authorinfo{* Denote equal contribution. Contact author: Mathias Unberath (unberath@jhu.edu).}

\begin{document} 
\maketitle

\begin{abstract}
Fully supervised deep learning (DL) models for surgical video segmentation have been shown to struggle with non-adversarial, real-world corruptions of image quality  including smoke, bleeding, and low illumination.
Foundation models for image segmentation,  such as the  segment anything model (SAM) that focuses on interactive prompt-based segmentation, move away from semantic classes and thus can be trained on larger and more diverse data, which offers outstanding zero-shot generalization with appropriate user prompts. 
%
%
Recently, building upon this success, SAM-2 has been proposed to further extend the zero-shot interactive segmentation capabilities from independent frame-by-frame to video segmentation.
In this paper, we present a first experimental study evaluating SAM-2's performance on surgical video data.
Leveraging the \textit{SegSTRONG-C} MICCAI EndoVIS 2024 sub-challenge dataset, we assess SAM-2's effectiveness on uncorrupted endoscopic sequences and evaluate its non-adversarial robustness on videos with corrupted image quality simulating smoke, bleeding, and low brightness conditions under various prompt strategies.
Our experiments demonstrate that SAM-2, in zero-shot manner, can achieve competitive or even superior performance compared to fully-supervised deep learning models on surgical video data, including under non-adversarial corruptions of image quality. 
Additionally, SAM-2 consistently outperforms the original SAM and its medical variants across all conditions.
Finally, frame-sparse prompting can consistently outperform frame-wise prompting for SAM-2, suggesting that allowing SAM-2 to leverage its temporal modeling capabilities leads to more coherent and accurate segmentation compared to frequent prompting.
\end{abstract}


\section{Introduction}
Accurate segmentation of surgical tools and anatomical structures in endoscopic videos is fundamental for computer-assisted interventions and surgical robotics applications \cite{hasan2021segmentation}, such as  instrument tracking, surgical scene understanding, and automated surgical skill assessment \cite{lavanchy2023preserving}. 
However, real-world clinical scenarios present challenges for robust segmentation, including visual corruptions like smoke, bleeding, specular reflections, and low illumination that can substantially alter the appearance of surgical scenes \cite{kitaguchi2022limited,rueckert2024methods}. 
These challenging conditions, which are commonly encountered during actual procedures, can degrade segmentation performance and potentially compromise patient safety if not properly addressed \cite{hayoz2023learning}.
While fully supervised deep learning (DL) models have achieved promising results on standard surgical video datasets under controlled conditions, they often struggle to generalize effectively to the diverse and dynamic environments of live surgeries \cite{yang2022weakly}, which highlights the need for more robust and adaptable segmentation approaches that can maintain high accuracy across a wide range of real-world surgical scenarios
Recently, foundation models for image segmentation, particularly the Segment Anything Model (SAM), have demonstrated zero-shot capabilities when provided with appropriate user prompts \cite{sam}. 
SAM introduced a paradigm shift towards interactive segmentation, where a single model can be flexibly applied to diverse segmentation tasks through prompting. 
However, direct application of SAM to medical imaging domains like surgical videos has revealed performance gaps compared to fully supervised DL models trained specifically on medical data \cite{medsam,sammed2d}.
Building upon SAM's success, the recently proposed SAM-2 extends these zero-shot interactive segmentation capabilities to video data \cite{sam2}. 
SAM-2 introduces architectural modifications to handle temporal information in video data, including a memory encoder and memory bank to maintain consistency across video frames.
These enhancements make SAM-2 a promising candidate for robust surgical video segmentation, as it can potentially capture the dynamic nature of surgical procedures.

In this paper, we present the first experimental study comprehensively evaluating SAM-2's performance on surgical video data. 
We leverage the \textit{SegSTRONG-C} dataset \cite{segstrong}, which provides mock endoscopic sequences both with and without simulated visual corruptions like smoke, bleeding, and low brightness to assess SAM-2's effectiveness on standard surgical videos as well as its robustness to challenging real-world conditions.
Our study compares SAM-2 against the original SAM, medical variants of SAM that process video as per-frame image segmentation, and state-of-the-art fully supervised DL models. 
We explore various prompting strategies to investigate how SAM-2 can best leverage its temporal modeling capabilities. 
Our experiments reveal SAM-2's strengths in maintaining temporal consistency across video frames and its adaptability to surgical scenarios, even outperforming fully supervised models in some cases.
The major contributions are three-fold.
Firstly, we conduct the first comprehensive evaluation of SAM-2 on surgical video data, including uncorrupted sequences and those with simulated visual corruptions.
Secondly, we provide an in-depth comparison of SAM-2 against other SAM variants and fully supervised models, highlighting the benefits of SAM-2's temporal modeling for surgical videos.
Finally, we explore various prompting strategies to optimize SAM-2's performance on surgical data, including frame-wise and frame-sparse prompting.

\section{Methods}

\paragraph{Preliminary of SAM-2}
SAM-2 \cite{sam2} is an extension of the original SAM \cite{sam}, designed to handle video segmentation tasks. 
It builds upon SAM's architecture with modifications to accommodate temporal information. 
Specifically, it consists of five main components, namely an image encoder, a prompt encoder, a mask decoder, a memory encoder, and a memory bank. 
The image encoder processes each video frame to generate a high-dimensional embedding, while the prompt encoder embed the user inputs such as point prompt.
The mask decoder combines the image embedding and prompt embedding to produce segmentation masks. 
The memory encoder compresses and encodes past frame information and predictions, which are then stored into the memory bank. 
SAM-2 introduces a memory bank that retains information about past predictions, allowing for temporal consistency across frames.

\paragraph{Evaluation Dataset and Evaluation Metrics}
We leverage the test sets of the \textit{SegSTRONG-C} from MICCAI EndoVIS 2024 sub-challenge to evaluate the robustness of SAM-2 against common surgical complications \cite{segstrong}.
This test dataset comprises $3$ mock endoscopic video sequences, each containing 300 frames for both left and right cameras, with corresponding binary segmentation masks for surgical tools in each frame.
Our evaluation encompasses five distinct conditions. 
The regular condition uses unmodified surgical video sequences, representing ideal conditions. 
To assess adaptability to varying anatomical contexts, we employ a background change (`BG Change') condition where videos feature altered background tissue. 
We also evaluate robustness under three simulated surgical complications, including bleeding, where artificial blood is introduced into the surgical field; smoke, mimicking the presence of surgical smoke during electrocautery; and low brightness, simulating sub-optimal lighting conditions. 
These three conditions enable the evaluation of SAM-2's ability to maintain accurate segmentations across challenging situations commonly encountered in real-world surgical environments.
We utilize the Dice Similarity Coefficient (`DSC') and Normalized Surface Distance (`NSD') as the evaluation metrics. 
These two metrics are calculated and averaged across all frames in each video sequence.

\paragraph{Prompting Strategies}
We consistently utilize point prompts across our experiments for both SAM-2 and compared SAMs. 
To ensure a fair comparison with the original SAM \cite{sam} and its medical variants (MedSAM \cite{medsam} and SAM-Med2D \cite{sammed2d}), we first employ frame-wise prompting. 
This approach consists of providing $K$ positive and $K$ negative clicks on the object of interest within each frame.
Positive clicks are sampled from the ground truth mask, with a bias towards the object's center and edges, while negative clicks are sampled from areas outside the ground truth mask.
To leverage SAM-2's temporal capabilities, we also explore frame-spare prompting strategies, where prompts are provided only for the first frame within a subsequent sequence of $N$ frames \textit{i}.\textit{e}., updating prompts periodically every $N$ frames.
For comparison, when using the original SAM in this frame-spare setting, we simulate a basic form of temporal consistency by leveraging predictions from the previous frame to prompt the next frame within the $N$-frame sequence.
It's worth noting that frame-wise prompting can be considered a special case of frame-sparse prompting where $N=1$.

\begin{table}[ht!]
\centering
\caption{Experiment results of SAM-2 with SAM \cite{sam}, its medical variants \cite{medsam,sammed2d}, and fully-supervised DL methods \cite{unet,deeplabv3,segformer}.
Results are reported for five conditions: Regular, Background Change (`BG Change'), Bleeding, Smoke, and Low Brightness.
Frame-wise prompting provides prompts every frame ($N=1$), while frame-sparse prompting provides prompts every 300 frames ($N=300$).
}\label{tab:result}
\resizebox{\textwidth}{!}{
\begin{tabular}{llcccccccccc}
\toprule
\multirow{2}{*}{Prompt Strategy} & \multirow{2}{*}{Models} & \multicolumn{5}{c}{NSD ($\uparrow$)} & \multicolumn{5}{c}{DSC ($\uparrow$)} \\
\cmidrule(lr){3-7} \cmidrule(lr){8-12}
& & Regular & BG Change & Bleeding & Smoke & Low Brightness & Regular & BG Change & Bleeding & Smoke & Low Brightness \\
\midrule
\multirow{3}{*}{Fully Supervised} & UNet \cite{unet} & 0.8888 & 0.7379 & 0.5677 & 0.5084 & 0.4390 & 0.9372 & 0.8878 & 0.7052 & 0.6603 & 0.5750 \\
& DeepLabv3+ \cite{deeplabv3} & 0.7941 & 0.6140 & 0.5629 & 0.4637 & 0.4000 & 0.8961 & 0.8102 & 0.6896 & 0.6538 & 0.5352 \\
& SegFormer \cite{segformer} & 0.8023 & 0.5962 & 0.5133 & 0.5266 & 0.4194 & 0.8993 & 0.7864 & 0.6802 & 0.6906 & 0.6145 \\
\midrule
\multirow{4}{*}{\makecell[l]{Frame-wise\\Point Prompt\\($N=1$)}} 
& SAM \cite{sam} & 0.6897 & 0.7419 & 0.5643 & 0.6260 & 0.2281 & 0.8496 & 0.8791 & 0.7681 & 0.8134 & 0.4088 \\
& MedSAM \cite{medsam} & 0.1611 & 0.1771 & 0.1636 & 0.1270 & 0.1418 &  0.2059 & 0.2511 & 0.2213 & 0.1408 & 0.2284\\
& SAM-Med2D \cite{sammed2d} & 0.4156 & 0.4508 & 0.3393 & 0.3546 & 0.2423 & 0.6307 & 0.6852 & 0.5302 & 0.5791 & 0.4132 \\
& SAM-2 \cite{sam2}  & 0.7315 & 0.8618 & 0.6545  & 0.6466  & 0.2396  & 0.8543  & 0.9168 & 0.8134  & 0.7927  & 0.4616 \\
\midrule
\multirow{3}{*}{\makecell[l]{Frame-wise\\BBox Prompt\\($N=1$)}} 
& SAM \cite{sam} & 0.6228 & 0.6272 & 0.5846 & 0.5884 & 0.2586 & 0.7952 & 0.7956 & 0.7496 & 0.7475 & 0.4846 \\
& MedSAM \cite{medsam} & 0.2423 & 0.3672 & 0.2347 & 0.2304 & 0.1577 & 0.4604 & 0.6209 & 0.4743 & 0.4582 & 0.4762\\
& SAM-Med2D \cite{sammed2d} & 0.3377 & 0.3398 & 0.2748 & 0.2891 & 0.2314  & 0.5055 & 0.4389 & 0.3448 & 0.4689 & 0.4143  \\
\midrule
\multirow{4}{*}{\makecell[l]{Frame-sparse\\Point Prompt\\($N=10$)}} 
& SAM \cite{sam} & 0.7396 & 0.6935 & 0.6852 & 0.7128 & 0.2890 & 0.8539 & 0.8307 & 0.8180 & 0.8394 & 0.5221 \\
& MedSAM \cite{medsam} & 0.1284 & 0.1270 & 0.1302 & 0.1185 & 0.1192 &  0.1779 & 0.2183 & 0.2108 & 0.1797 & 0.1806\\
& SAM-Med2D \cite{sammed2d} & 0.2787 & 0.3092 & 0.2440 & 0.2624 & 0.2671 & 0.5203 & 0.6003 & 0.4503 & 0.5006 & 0.4627 \\
& SAM-2 \cite{sam2}  & 0.8182 & 0.9023 & 0.7982  & 0.7548  & 0.2570  & 0.8935  & 0.9267 & 0.8714  & 0.8484  & 0.4335 \\
\midrule
\multirow{3}{*}{\makecell[l]{Frame-sparse\\Bbox Prompt\\($N=10$)}} 
& SAM \cite{sam} & 0.6108 & 0.6259 & 0.5307 & 0.5458 & 0.2554 & 0.7721 & 0.7766 & 0.7125 & 0.7230 & 0.4505 \\
& MedSAM \cite{medsam} & 0.1963 & 0.3372 & 0.1881 & 0.1973 & 0.1665 & 0.4519 & 0.5902 & 0.4606 & 0.4560 & 0.4630\\
& SAM-Med2D \cite{sammed2d} & 0.2687 & 0.2686 & 0.2337 & 0.2316 & 0.2313  & 0.4473 & 0.3862 & 0.3004 &  0.4033 & 0.4062  \\
\midrule
\multirow{4}{*}{\makecell[l]{Frame-sparse\\Point Prompt\\($N=300$)}} 
& SAM \cite{sam} & 0.4889 & 0.5327 & 0.4180 & 0.4580 & 0.2321 &  0.6026 & 0.6132 & 0.5458 & 0.5746 & 0.3697 \\
& MedSAM \cite{medsam} & 0.1218 & 0.1237 & 0.1225 & 0.1081 & 0.1041 & 0.1214 & 0.1454 & 0.1276 & 0.1102 & 0.1249 \\
& SAM-Med2D \cite{sammed2d}& 0.3606 & 0.3745 & 0.3003 & 0.3001 & 0.2370 & 0.5428 & 0.5684 & 0.4400 & 0.4914 & 0.3919\\
& SAM-2 \cite{sam2}  & 0.8479 & 0.9186 & 0.8002  & 0.7808  & 0.2882  & 0.9096  & 0.9325 & 0.8628  & 0.8660  & 0.4373 \\
\midrule
\multirow{3}{*}{\makecell[l]{Frame-sparse\\Bbox Prompt\\($N=300$)}} 
 & SAM \cite{sam} & 0.4687 & 0.5441 & 0.4748 & 0.4391 & 0.2591 &  0.6157 & 0.6403 & 0.6321 & 0.5954 & 0.4253 \\
 & MedSAM \cite{medsam} & 0.2276 & 0.3591 & 0.2187 & 0.2250& 0.2158 & 0.4540 & 0.6350 & 0.4634 &  0.4625 &  0.4605 \\
 & SAM-Med2D \cite{sammed2d}& 0.1221 & 0.1422 & 0.1149 &  0.1522 & 0.1437  & 0.3447 & 0.3067 &  0.3208 & 0.3211 &  0.3248 \\
\bottomrule
\end{tabular}
}
\end{table}

\section{Experiments}
\begin{figure}[ht!]
\centering
\centerline{\includegraphics[width=0.9\linewidth]{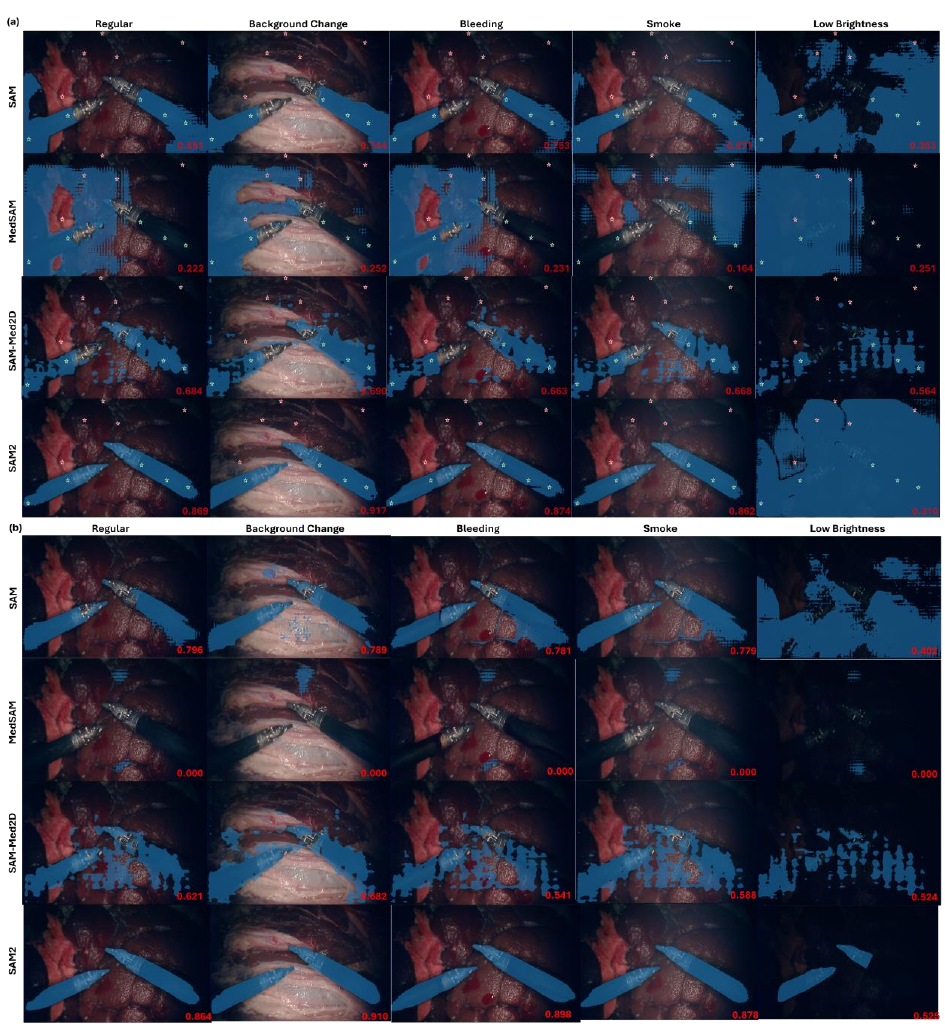}
}
\caption{The segmentation results of various SAM on the last frame with different prompting strategy. (a) Frame-wise prompting. (b) Frame-spare prompting. 
Green and red starts denote the positive and negative point prompts respectively.}
\label{fig:result}
\end{figure}

Table \ref{tab:result} presents a comprehensive comparison of SAM-2's performance against SAMs and fully-supervised DL models across five surgical video conditions. 
Segmentation results are visualized in Fig.~\ref{fig:result}.
We set $K=5$ in this study.

\paragraph{Performance on Regular Videos} 
SAM-2 demonstrates competitive or superior performance compared to fully-supervised DL models across most conditions. 
On regular surgical videos, SAM-2 demonstrates competitive performance compared to them. 
With frame-wise point prompting (\textit{i}.\textit{e}., $N=1$), SAM-2 achieves a DSC of $0.8543$ in a zero-shot manner, which is comparable to the best-performing fully-supervised model, UNet with DSC of $0.9372$.
Notably, when using frame-sparse point prompting ($N=300$), SAM-2's performance improves, reaching a DSC of $0.9096$, which is very close to UNet's performance and can outperform DeepLabv3+ and SegFormer.

\paragraph{Robustness on Corrupted Videos}
SAM-2 exhibits good robustness across various simulated surgical complications.
For example, in the background change condition, SAM-2 outperforms all other models, achieving the highest DSC of $0.9325$ with frame-sparse prompting, surpassing even the best fully-supervised DL model \textit{i}.\textit{e}., UNet ($0.8878$).
For bleeding scenarios, SAM-2 demonstrates superior performance with a DSC of $0.8628$ using frame-sparse prompting, outperforming all fully-supervised DL models, original SAM and its medical variants. 
In smoky conditions, SAM-2 maintains high performance with a DSC of $0.8628$ with frame-sparse prompting, again substantially outperforming all other models. 
While all models struggle in low brightness conditions, SAM-2 with frame-sparse prompting achieves a DSC of $0.4373$ and outperforms all other SAM models.

\paragraph{Comparison with SAMs}
Under the frame-sparse point prompt setting ($N=300$), SAM-2 consistently outperforms the original SAM and its medical variants across all five conditions. 
The performance gap is particularly notable in challenging three conditions. 
For instance, in the bleeding scenario, SAM-2 achieves a DSC of $0.8628$, surpassing SAM ($0.5458$), MedSAM ($0.1276$), and SAM-Med2D ($0.4400$).
Similarly, in smoky conditions, SAM-2 maintains high DSC performance ($0.8660$) compared to SAM ($0.5746$), MedSAM ($0.1102$), and SAM-Med2D ($0.5746$).
Even in regular conditions, SAM-2 outperforms its counterparts by a large margin.
The original SAM performs reasonably well with frame-wise prompting but degrades significantly with frame-sparse prompting, highlighting SAM-2's better temporal modeling capabilities.

\paragraph{Impact of Prompting Strategy}
Frame-sparse point prompting ($N=300$) consistently improves SAM-2's performance compared to frame-wise point prompting ($N=1$) across all conditions. 
This improvement is substantial, with DSC increases of 6.47\%, 1.71\%, 6.07\%, 9.25\%, and -5.26\% for regular, background change, bleeding, smoke, and low brightness conditions, respectively.
These results suggest that using fewer prompts can actually lead to better performance for SAM-2 on the surgical videos from \textit{SegSTRONG-C} dataset \cite{segstrong}.
This counter-intuitive finding can be attributed to the contribution of SAM-2's memory encoder and bank in leveraging temporal information.
In contrast, frame-wise point prompting may not consistently select optimal prompts, potentially introducing noise or inconsistencies in the segmentation process.
By allowing SAM-2 to rely more on its temporal modeling capabilities, frame-sparse prompting enables the model to maintain coherence across frames while reducing the impact of suboptimal individual prompts.
Moreover, SAM \cite{sam} and SAM-Med2D \cite{sammed2d} show decreased performance with frame-sparse prompting, underscoring the unique advantage of SAM-2's architecture in handling temporal data.

\section{Conclusion}
This experimental study provides the first comprehensive evaluation of SAM-2's performance on surgical video data, demonstrating its potential as a robust and adaptable solution for surgical tool segmentation.
Our experiments on the \textit{SegSTRONG-C} dataset reveal that SAM-2 achieves competitive or superior performance compared to fully-supervised deep learning models across various surgical conditions, including uncorrupted videos and those with simulated visual corruptions.
SAM-2 consistently outperforms the original SAM and its medical variants, particularly in challenging scenarios, highlighting the benefits of its temporal modeling capabilities for surgical videos. 
Notably, frame-sparse prompting enhances SAM-2's performance compared to frame-wise prompting, suggesting that leveraging temporal information can lead to more coherent and accurate segmentation.
These findings indicate that SAM-2 represents a promising zero-shot method in surgical video segmentation.

\bibliography{report} 
\bibliographystyle{spiebib} 

\end{document}